\documentclass{ws-procs975x65}

\def\beq{\begin{equation}}
\def\eeq{\end{equation}}

\begin{document}

\title{Exploring a missing link between peculiar, sub- and super-Chandrasekhar type Ia supernovae by modifying Einstein's gravity}
\author{Banibrata Mukhopadhyay}

\address{Department of Physics, Indian Institute of Science, Bangalore 560012\\
E-mail: bm@physics.iisc.ernet.in\\
}

%

\begin{abstract}
Observations of several peculiar, under- and over-luminous type Ia supernovae (SNeIa) argue 
for exploding masses widely different from the Chandrasekhar-limit. We explore the modification to Einstein's 
gravity in white dwarfs for the first time in the literature, which shows that depending on the (density dependent)
modified gravity parameter $\alpha$, chosen for the present purpose of representation, limiting mass
of white dwarfs could be significantly sub- as well as super-Chandrasekhar. Hence, this unifies 
the apparently disjoint classes
of SNeIa, establishing the importance of modified Einstein's gravity in white dwarfs.
Our discovery questions both the global validity of Einstein's gravity
and the uniqueness of Chandrasekhar's limit.
\end{abstract}


\bodymatter


\section{Introduction}\label{intro}


%
%
%

Since last few years, we have been exploring physics behind peculiar type~Ia supernovae (SNeIa), which 
are highly over-luminous, and the possible existence of super-Chandrasekhar white dwarfs. Our initiation 
has brought the topic of super-Chandrasekhar white dwarfs in limelight, with so many papers published following us.

SNeIa are believed to be triggered from the
violent thermonuclear explosion of a carbon-oxygen white dwarf
on approaching its mass the Chandrasekhar limit of
$1.44M_\odot$ \cite{chandra35}. 
SNIa is used as a standard candle in understanding
the expansion history of the universe \cite{perl99}.
Nevertheless, some of these SNeIa are highly over-luminous, e.g. SN 2003fg, SN 2006gz, SN 2007if, SN 2009dc
\cite{howel,scalzo},
and some others are highly under-luminous, e.g. SN 1991bg, SN 1997cn, SN 1998de, SN 1999by
\cite{1991bg,taub2008}.
The luminosity of the former group (super-SNeIa) implies the existence of highly super-Chandrasekhar progenitor
white dwarfs with mass $2.1-2.8M_\odot$ \cite{howel,scalzo}.
While, the latter group (sub-SNeIa) predicts the progenitor mass to be as low
as $ \sim M_\odot$ \cite{1991bg}.

While we argued, in a series of papers, that highly magnetized white dwarfs could be
as massive as inferred from the above super-SNeIa observations \cite{prd,prll,jcap14}, they
are unable to explain the sub-SNeIa.
All the previous models proposed to describe them entail caveats.
For example, although numerical simulations of the merger of two sub-Chandrasekhar white dwarfs
reproduce the sub-SNeIa, the underlying simulated light-curves fade
slower than that suggested by observations.

Nonetheless, a major concern is a large number of 
of models required to explain apparently the same phenomena, i.e., triggering of thermonuclear explosions
in white dwarfs. Why there are mutually uncorrelated sub- and super-SNeIa in nature?
This is where the proposal of modifying general relativity steps in into the context of white dwarfs. We will show that
modified general relativity unifies the sub-classes of SNeIa by a single underlying theory, hence serve as a missing link.

\section{Basic equations and formalism}\label{formal}

Let us start with the 4-dimensional action as \cite{livrel}
\begin{equation}
S = \int \left[\frac{1}{16\pi} f(R) + {\cal L}_M \right] \sqrt{-g}~ d^4 x , 
\label{action}
\end{equation}
where $g$ is the determinant of the spacetime metric $g_{\mu\nu}$, $d^4 x$ the 4-dimensional 
volume element,  
${\cal L}_M$ the Lagrangian density of the matter field, 
$R$ the scalar curvature defined as $R=g^{\mu\nu}R_{\mu\nu}$, 
where $R_{\mu\nu}$ is the Ricci tensor and $f$ is an 
arbitrary function of $R$; in general relativity, $f(R)=R$. 

For the present purpose, we consider the simplistic Starobinsky model 
\cite{staro} defined as $f(R)=R+\alpha R^2$, when $\alpha$ is a constant. 
However, similar effects could also be obtained in other, physically more sophisticated, theories, where
$\alpha$ (or effective-$\alpha$) is varying (e.g., with density). Now, on extremizing the action Eq. (\ref{action})
for Starobinsky's model, 
one obtains the modified field equation of the form
\begin{equation}
G_{\mu\nu} + \alpha \left[2 R G_{\mu\nu} + \frac{1}{2} R^2 g_{\mu\nu} - 2(\nabla_\mu \nabla_\nu - g_{\mu\nu}\Box)R \right] = 8\pi T_{\mu\nu},
\label{modfld}
\end{equation}
where $T_{\mu\nu}$ contains only the matter field (non-magnetic star),
$G_{\mu\nu}$ is Einstein's field tensor, 
$\nabla_\mu$ and $\nabla_\nu$ are the covariant derivatives and $\Box=\nabla_\mu\nabla^\mu$ (see Refs.~\refcite{jcap15b,curr}, for details).

For the present purpose, we seek perturbative solutions of Eq. (\ref{modfld}) (see, e.g., Ref.~\refcite{capo}), 
such that $\alpha R\ll 1$. Furthermore, we consider the hydrostatic equilibrium condition
so that $g_{\nu r}\nabla_\mu T^{\mu\nu}=0$, 
with zero velocity. Hence, we
obtain the differential equations for mass $M_\alpha (r)$, pressure $P_\alpha (r)$
(or density $\rho_\alpha (r)$)
and gravitational potential $\phi_\alpha (r)$, of spherically symmetric white dwarfs
(which is basically the set of {\it modified} Tolman-Oppenheimer-Volkoff (TOV) equations). 
For $\alpha=0$, these  equations reduce to TOV equations in general relativity.

As the white dwarf is assumed to be nonmagnetized, we consider EoS, as obtained 
by Chandrasekhar \cite{chandra35} at extremely low and high densities, 
$P_0=K\rho_0^{1+(1/n)}$, where $P$ and $\rho$ of Ref.~\refcite{chandra35}
are replaced by $P_0$ and $\rho_0$ respectively ($\alpha=0$: general relativity) in the spirit of perturbative approach. 
Here, $n$ is the polytropic index and $K$ a dimensional constant. 
The boundary conditions are: $M_\alpha (0)=0$ and $\rho_\alpha (0) =\rho_c$ (central mass and density respectively). 
Note that 
by varying $\rho_c$ from $2\times 10^5$ gm/cc to $10^{11}$ gm/cc, we construct 
the mass-radius relation of white dwarfs. 

\section{Results}

We show in Figs. 1(a) and (b) 
that all three $M_\alpha-\rho_c$ curves for $\alpha>0$ overlap with the $\alpha=0$ curve in the low density region. 
However, as $\alpha$ increases, the region of overlap decreases, receding to a lower $\rho_c$ region. 
At $\rho_c \gtrsim 10^8,~4\times 10^7$ and $2\times 10^6$ gm/cc, modified general relativity
effects become important and visible 
for $\alpha = 2\times 10^{13}~ {\rm cm^2}$, $8\times 10^{13}~ {\rm cm^2}$ and 
$10^{15}~{\rm cm^2}$ respectively. At a fixed $\alpha$, with the increase of $\rho_c$, first $M_\alpha$ 
increases, then by reaching a maximum value starts decreasing, like the $\alpha=0$ (general relativity) case. 
The maximum mass $M_{\rm max}$ decreases with the 
increasing $\alpha$ and for $\alpha=10^{15}~{\rm cm}^2$ it is as low as $0.81M_\odot$ (highly sub-Chandrasekhar).
This argues that modified general relativity has a tremendous impact on white dwarfs.
In fact, $0.81\lesssim M_{\rm max}/M_\odot\lesssim 1.31$ for all the chosen $\alpha>0$. 
This is a remarkable finding since it establishes that even if $\rho_c$s for 
these sub-Chandrasekhar white dwarfs are lower than the conventional value at which SNeIa are usually 
triggered, an attempt 
to increase the mass beyond $M_{\rm max}$ with the increase of $\rho_c$ will lead to a gravitational instability. 
Subsequently, this presumably will be leading to a 
runaway thermonuclear reaction, provided the core temperature increases sufficiently due to collapse. 
Occurrence of such thermonuclear runway reactions, triggered at a low density as $10^6$ gm/cc,
has already been demonstrated \cite{runaway}. Thus, once $M_{\rm max}$ is approached for white dwarfs with 
$\alpha>0$, a SNIa is expected to trigger just 
like in the $\alpha=0$ case, explaining the sub-SNeIa \cite{1991bg,taub2008}, 
like SN 1991bg mentioned above. 


\begin{figure}[h]
\begin{center}
\includegraphics[angle=0,width=15.5cm]{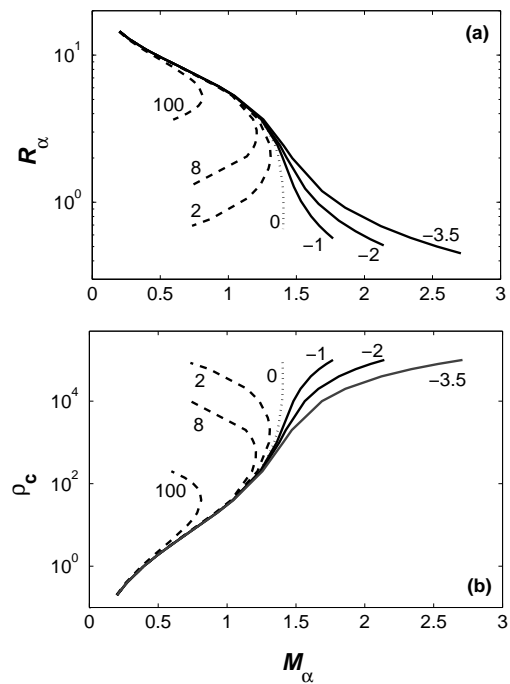}
\caption{Unification of under-luminous and over-luminous SNeIa:
(a) mass-radius relations, (b) variation of $\rho_c$ with $M_\alpha$.
The numbers adjacent to the various lines denote 
$\alpha/(10^{13}~{\rm cm^2})$. 
$\rho_c$, $M_\alpha$ and $R_\alpha$ are in units of $10^6$ gm/cc, $M_\odot$ and 1000 km respectively.
  }
\label{fig}
\end{center}
\end{figure}

Figure 1(b) shows that for $\rho_c>10^8$ gm/cc with $\alpha<0$, the 
$M_\alpha-\rho_c$ curves deviate from the general relativity curve due to modified general relativity effects. 
Note that $M_{\rm max}$ for all the three cases corresponds to $\rho_c=10^{11}$ gm/cc, 
what upper-limit is chosen to avoid possible neutron-drip.
Interestingly, all values of $M_{\rm max}$, lying in $1.8-2.7M_\odot$, are highly super-Chandrasekhar.
Thus, while the general relativity effect is very small (but non-negligible), modified general relativity effect could
lead to $\sim 100\%$ increase in the limiting mass.
The corresponding values of $\rho_c$ are large enough, i.e. larger than $\rho_c$ corresponding to $M_{\rm max}$ of 
$\alpha=0$ case, to initiate thermonuclear reactions,
whereas the respective 
core temperatures are expected to be similar. This 
explains the entire range of the observed super-SNeIa mentioned above \cite{howel,scalzo},
assuming the furthermore gaining mass above $M_{\rm max}$ leads to SNeIa.


Table 1 ensures the validity of perturbation approach of the solutions, where
we solve the modified TOV equations only up to ${\cal O} (\alpha)$. Since the product 
$\alpha R$ is first order in $\alpha$, we replace $R$ in it by the zero-th order Ricci scalar 
$R^{(0)} = 8\pi(\rho^{(0)} - 3P^{(0)})$, i.e. Ricci scalar in general relativity ($\alpha=0$). 
For the perturbative validity of the entire solution, 
$|\alpha R^{(0)}|_{\rm max} \ll 1$ should satisfy.
Next, we consider $g_{tt}^{(0)}/g_{tt}$ and $g_{rr}^{(0)}/g_{rr}$ (ratios of $g_{\mu\nu}$-s 
in general relativity and those in modified general relativity up to ${\cal O} (\alpha)$),
which should be close to unity for the validity of perturbative method \cite{orelana}. Hence,
$|1-g_{tt}^{(0)}/g_{tt}|_{\rm max} \ll 1$ and 
$|1-g_{rr}^{(0)}/g_{rr}|_{\rm max} \ll 1$ should both satisfy.
Table 1 indeed shows that all three measures quantifying the validity of perturbative are at least $2-3$ 
orders of magnitude smaller than 1. 

\begin{table}
\tbl{Measure of validity of perturbative solutions corresponding to $M_{\rm max}$ in Fig. \ref{fig}.}
{\begin{tabular}{@{}cccc@{}}
\toprule
$\alpha/(10^{13}~{\rm cm}^2)$ & $|\alpha R^{(0)}|_{\rm max}$ & $|1-g^{(0)}_{tt}/g_{tt}|_{\rm max}$ & $|1-g^{(0)}_{rr}/g_{rr}|_{\rm max}$\\
\hline
2 & $7.4\times 10^{-5}$  & $6.8\times 10^{-5}$ & $2.0\times 10^{-4}$ \\
8 & $7.4\times 10^{-5}$ & $6.8\times 10^{-5}$ & $2.0\times 10^{-4}$ \\
100 & $7.4\times 10^{-5}$ & $6.9\times 10^{-5}$  & $2.0\times 10^{-4}$ \\
%
\hline
-1 & 0.00184 & 0.0016 & 0.0052 \\ 
-2 &  0.00369 & 0.0031 & 0.0108 \\
-3.5 & 0.00646 & 0.0052 & 0.0199 \\ \hline
\botrule
\end{tabular}}
\label{aba:tbl1}
\end{table}

\section{Possible chameleon-like effect for density dependent model parameter}

we now justify that the effects of modified general relativity based on a more sophisticated calculation, invoking
an (effective) $\alpha$ varying explicitly with density (and effectively becoming negative), are 
likely to converge to those described above.
Note that even though $\alpha$ is assumed to be constant within
individual white dwarfs here, there is indeed an implicit
dependence of $\alpha$ on $\rho_c$, clearly shown in Fig. \ref{fig}(b) for limiting mass white dwarfs presumably leading to
SNeIa. This indicates the existence
of an underlying chameleon effect, which trend is expected to emerge self-consistently in a varying-$\alpha$ theory.

Let us consider a possible situation where $\alpha$ is varying explicitly with density and 
try to relate it with the results presented above. Note the fact that the super-SNeIa occur mostly in young stellar 
populations consisting of massive stars (see, e.g., Ref.~\refcite{howel}),
while the sub-SNeIa occur in old stellar
populations consisting of low mass stars (see, e.g., Ref.~\refcite{gonz}). The massive stars with
higher densities are likely to give rise to
super-Chandrasekhar white dwarfs on collapse,
which, on gaining mass, would subsequently explode to produce super-SNeIa. The low mass
stars with lower densities would be expected to give rise to sub-Chandrasekhar white dwarfs on collapse,
which furthermore would probably end with sub-SNeIa. Now, let us assume
$\alpha$ to be depending on density in such a way that there are two terms ---
one with negative sign dominates at higher densities and the other with positive sign dominates at lower densities. Hence,
when a massive, high density star collapses, it results in similar to our $\alpha < 0$ cases;
while a low mass, low density star collapse leads to results
like our $\alpha > 0$ cases. Thus, the same functional
form of $\alpha$ could lead to both super- and sub-Chandrasekhar
limiting mass white dwarfs, respectively, depending on their densities. Of course, the final
mass of the white dwarf would depend on several factors,
such as, $\rho_c$ and the density gradient in
the parent star, etc. Interestingly, this description of 
density dependent $\alpha$ is essentially equivalent to invoking
a so-called ``chameleon-$f(R)$ theory", which can pass solar
system tests of gravity (see, e.g., Ref.~\refcite{cham2}).
This is so because, once $\alpha$ is a function of density, it
is a function of $R$. Hence, introduction of a
density (and hence $R$) dependent $\alpha$ is equivalent of choosing an
appropriate (more complicated) $f(R)$ model of gravity.
Therefore, a more self-consistent variation of $\alpha$
with density does not invalidate the results of the
constant-$\alpha$ cases, rather is expected to complement the 
picture.

We must mention that the orders of magnitude of $\alpha$ are different
between typical white dwarfs ($\alpha_{\rm WD} \sim 10^{13}$ ${\rm cm^2}$, as used above) and neutron stars
($\alpha_{\rm NS} \sim 10^9$ ${\rm cm^2}$, e.g. \cite{eksi,capo}). This again argues for the fact
that there is an underlying chameleon effect which causes
$\alpha$ to be different in different density regimes. 
Now, the quantity $\alpha R$ would have a
similar value in both neutron stars and white dwarfs in the
perturbative regime. Hence, due to their higher curvature and density, neutron stars
will harbor a smaller value of $\alpha$ compared to white dwarfs. Roughly, neutron
stars are $10^4$ times denser than white dwarfs and, therefore,
$\alpha_{\rm NS}$ is $10^4$ times smaller than $\alpha_{\rm WD}$.
We also emphasize that the current work is an initiation
of the exploration of the effects of modified gravity in
white dwarfs, based on the motivation to explain observations
of peculiar SNeIa. Now, one has to polish the model step by step.

\section{Summary}\label{summ}

Based on a specific type of modified Einstein's gravity, namely simple Starobinsky $f(R)$-model, 
we show that modifications to general relativity
are indispensable in white dwarfs, in particular to explain observed data related to their limiting mass. 
It remarkably explains and unifies a wide range of SNeIa for which general relativity is insufficient.
Although the present study is based perturbative method, this is indeed useful
as then we have a handle on $\alpha$ characterizing our model, which has an upper bound from astrophysical observations \cite{nah}.
Hence, depending on the magnitude and sign of $\alpha$, we obtain both
highly super-Chandrasekhar and highly sub-Chandrasekhar 
limiting mass white dwarfs, which furthermore help to establish them as progenitors of the peculiar super- and
sub-SNeIa, respectively.
Thus, a single underlying theory, i.e. an $f(R)$-model,
unifies the two apparently, puzzling, disjoint sub-classes of SNeIa, hence serves as a missing link.
Our discovery raises two fundamental
questions. Is the Chandrasekhar limit unique? Is Einstein’s gravity the ultimate theory for
understanding astronomical phenomena? Both the answers appear to be no!

\section*{Acknowledgments}
The author thanks Upasana Das for discussion.


\begin{thebibliography}{0}

\bibitem{chandra35} S. Chandrasekhar, {\em MNRAS} {\bf 95}, 207 (1935).

\bibitem{perl99}
S. Perlmutter, et al., {\em Astrophys. J.} {\bf 517}, 565 (1999).

\bibitem{howel}
D.A. Howell, et al., {\em Nature} {\bf 443}, 308 (2006).

\bibitem{scalzo}
R.A. Scalzo, et al., {\em Astrophys. J.} {\bf 713}, 1073 (2010).

\bibitem{1991bg}
A.V. Filippenko, et al., {\em Astron. J.} {\bf 104}, 1543  (1992).

\bibitem{taub2008}
S. Taubenberger, et al., {\em MNRAS} {\bf 385}, 75 (2008).

\bibitem{prd} U. Das and B. Mukhopadhyay, {\em Phys. Rev. D} {\bf 86}, 042001 (2012).

\bibitem{prll} U. Das and B. Mukhopadhyay, {\em Phys. Rev. Lett.} {\bf 110}, 071102 (2013).

\bibitem{jcap14} U. Das and B. Mukhopadhyay, {\em JCAP} {\bf 06}, 050 (2014).

\bibitem{livrel} A. de Felice and S. Tsujikawa, {\em Liv. Rev. Rel.} {\bf 13}, 3 (2010).

\bibitem{staro} A.A. Starobinsky, {\em Phys. Lett. B} {\bf 91}, 99 (1980).

\bibitem{jcap15b} U. Das and B. Mukhopadhyay, {\em JCAP} {\bf 05}, 045 (2015).

\bibitem{curr} B. Mukhopadhyay, {\em Curr. Sc.} {\bf 109}, 2250 (2015).

\bibitem{capo} A.V. Astashenok, S. Capozziello and S.D. Odintsov, {\em JCAP} {\bf 12}, 040 (2013).

\bibitem{runaway}
I.R. Seitenzahl, C.A. Meakin, D.M. Townsley, D.Q. Lamb and J.W. Truran, {\em Astrophys. J} {\bf 696}, 515 (2009).

\bibitem{orelana}
M. Orellana, F. Garc\'{i}a, F.A. Teppa Pannia and G.E. Romero, {\em Gen. Rel. Grav.} {\bf 45}, 771 (2013).

\bibitem{gonz}
S. Gonz\'{a}lez-Gait\'{a}n, et al., {\em Astrophys. J} {\bf 727}, 107 (2011).

\bibitem{cham2} T. Faulkner, M. Tegmark, E.F. Bunn and Y. Mao, {\em Phys. Rev. D} {\bf 76}, 063505 (2007).

\bibitem{eksi} S. Arapo\u{g}lu, C. Deliduman and K.Y. Ek\c{s}i, {\em JCAP} {\bf 7}, 020 (2011).

\bibitem{nah} J. N\"af and P. Jetzer, {\em Phys. Rev. D} {\bf 81}, 104003 (2010).







\end{thebibliography}
\end{document}